\documentclass[%
 reprint,
showpacs,preprintnumbers,
 amsmath,amssymb,
 aps,
floatfix,
]{revtex4-1}

\usepackage{graphicx}
\usepackage{dcolumn}
\usepackage{bm}

\begin{document}


\title{Novel Search for TeV-Initiated Pair Cascades in the Intergalactic Medium}

\author{Wenlei Chen}
\email{wenleichen@wustl.edu}
\affiliation{
	Department of Physics and McDonnell Center for the Space Sciences, Washington University in Saint Louis, MO 63130, USA.\\
    School of Physics and Astronomy, University of Minnesota, Minneapolis, MN 55455, USA
}
\author{Manel Errando}
\author{James H. Buckley}
\author{Francesc Ferrer}
\affiliation{
	Department of Physics and McDonnell Center for the Space Sciences, Washington University in Saint Louis, MO 63130, USA.\\
}

\date{\today}

\begin{abstract}
TeV emission from blazars can be used to probe the intergalactic magnetic fields and measure their intensity, coherence length, and helicity. Intergalactic magnetic fields deflect the electron-positron pairs produced by very-high-energy gamma-rays from blazars, resulting in broadened beams of cascade gamma-rays developing along the projected direction of the blazar jet. We present an analysis that uses for the first time the jet orientation of 12 high-synchrotron-peaked (HSP) BL~Lac-type blazars from VLBA radio observations to search for signatures of pair haloes in the Fermi-LAT data. Our search improves the sensitivity of previous studies by taking the asymmetry of the pair haloes into account, increasing the signal to noise and reducing the possibility of systematics in determining the point spread function. Although there is no significant detection, a hint for an offset halo with a global significance of 2$\sigma$ is found in the $30-300$ GeV energy range, corresponding to an intergalactic magnetic field with $\sim 10^{-15}$ Gauss, consistent with the inferred field from a prior study using stacked HSP BL Lac objects.
\end{abstract}

\maketitle


\section{\label{sec:introduction}Introduction}

Cosmological magnetic fields play a key role in the formation and evolution of astrophysical structures on multiple scales, but little is known about their strength or origin. Weak seed fields generated in the early Universe might persist in their initial form effectively frozen into the intergalactic medium. Hence, probing the intergalactic magnetic field (IGMF) enables studies of primordial magnetogenesis in the early Universe, and has the potential to provide new constraints on long-standing open questions in cosmology and astrophysics.

The IGMF in large-scale voids is too weak to be directly measured using observations of the Zeeman effect and Faraday rotation in the emission from distant sources \cite{Han2017,Subramanian2016}. However, indirect methods provide constraints on the IGMF in a wide range of spatial and temporal scales. The non-detection of the effects of magnetic fields on the cosmic microwave background (CMB) sets an upper limit to the strength of the present-day IGMF at a few nano-Gauss ($\mathrm{nG}$) level at $\sim 1\ \mathrm{Mpc}$ scales \cite{Planck2016}.

Pair halos around distant active galactic nuclei (AGNs) can also provide constraints on very weak IGMFs. In intergalactic space, weak magnetic fields deflect relativistic electron-positron pairs that are produced by very-high-energy (VHE, $> 100$ GeV) $\gamma$-rays interacting with the diffuse infrared extragalactic background light (EBL). These pairs upscatter CMB photons to GeV energies, leading to an offset, extended $\gamma$-ray halo around AGN that could be detected by $\gamma$-ray telescopes \cite{Aharonian1994}. By modeling the intrinsic TeV spectra of blazars and adopting models of the EBL and the CMB, the cascade emission from a few individual blazars has been studied, providing constraints and tentative measurements of the present-day IGMF strength in the range $\sim10^{-20}-10^{-14}$ G \cite{Murase2008,Neronov2010a,Essey2011,Arlen2014,Tanaka2014}. Detailed searches for pair halos in data from the Fermi Large Area Telescope (Fermi-LAT) using the LAT's on-orbit point spread function (PSF) on a stacked sample of AGN have found no significant evidence of extended halos \cite{Neronov2011,Ackermann2013}. More recently, a similar analysis using a sample of 24 low-redshift high-synchrotron-peaked BL Lacs (HBLs) revealed a hint of pair-halo emission at $\sim 1$ GeV, consistent with a present-day IGMF strength $\sim 10^{-17}-10^{-15}$ G \cite{Chen2015a}. Another set of recent studies have found signatures of non-vanishing parity-odd correlators from $\gamma$-ray arrival directions observed by Fermi-LAT, which could be explained by a left-handed helicity of the IGMF with strength $\sim 10^{-14}$ G \cite{Tashiro2013,Tashiro2014,Tashiro2015,Chen2015b}. However, due to large uncertainties from both observational data and theoretical parameters, a definite detection of pair cascades and other pair-cooling processes (e.g., plasma instabilities in the intergalactic medium \cite{Broderick2012,Schlickeiser2012,Miniati2013,Sironi2014,Kempf2016}) is still lacking.

A good understanding of the instrumental PSF plays a crucial role in the direct search for pair halos using Fermi-LAT data. Neglecting effects such as uneven exposures or nearby sources, the Fermi-LAT PSF is azimuthally symmetric around a point source. However, pair cascades develop along the direction of the blazar jet axis \cite{Neronov2010b,Arlen2014}, leading to an asymmetric pair halo broadened along the projected direction of the jet. Direct evidence of pair cascades in the intergalactic medium can be obtained if such an asymmetric signature of offset halos is statistically distinguishable from the symmetric PSF. Radio interferometry can resolve AGN jets with angular resolution down to millarcsecond scales, which in some nearby sources can clearly indicate the jet orientation projected in the observational plane. In this study, we present a joint analysis of the Fermi-LAT data from 12 high-synchrotron-peaked (HSP) or intermediate-synchrotron-peaked (ISP) blazars with well-determined jet orientation from radio observations. A Monte Carlo (MC) model of pair-halo images is used to statistically test the Fermi-LAT data in a parameter space that includes the IGMF strength.

\begin{table}[tbp]
	\centering
	\caption{List of the selected sources.}
	\vspace{0.2cm}
	\begin{tabular}{ ccccc }
		\hline
		Blazar name & Redshift & ~~~$\overline{p.a.}$ ($^\circ$)~~~ & $\sigma_{p.a.}$ ($^\circ$) \\
		\hline
		4C +67.04 & 0.29 & 301.4 & 7.6 \\
		OC $-230.4$ & 0.56 & 326.4 & 7.7 \\
		MG1 J021114+1051 & 0.2 & 67.5 & 16.0 \\
		3C 66A & 0.444 & 181.7 & 8.8 \\
		PKS 0301$-243$ & 0.2657 & 232.2 & 4.4 \\
		S5 0716+71 & 0.127 & 15.5 & 5.1 \\
		1ES 1011+496 & 0.212 & 265.1 & 13.8 \\
		7C 1055+5644 & 0.143 & 261.9 & 17.2 \\
		Mrk 421 & 0.0308 & 330.0 & 1.2 \\
		OQ 240 & 0.6 & 143.4 & 2.1 \\
		Mrk 501 & 0.0337 & 144.6 & 16.4 \\
		1ES 2344+514 & 0.044 & 139.1 & 2.1 \\
		\hline
	\end{tabular}
	\label{Tab:1}     
\end{table}

\section{\label{sec:source selection}Source selection}

In this study we select GeV-detected HSP and ISP blazars with measured jet orientations from the MOJAVE program. We only include sources with known redshifts in order to perform sufficiently accurate pair-halo modeling to allow a robust statistical search for halo emission \cite{note1}.

\begin{figure}[tbp]
	\centering
	\includegraphics[width=7cm, angle=0]{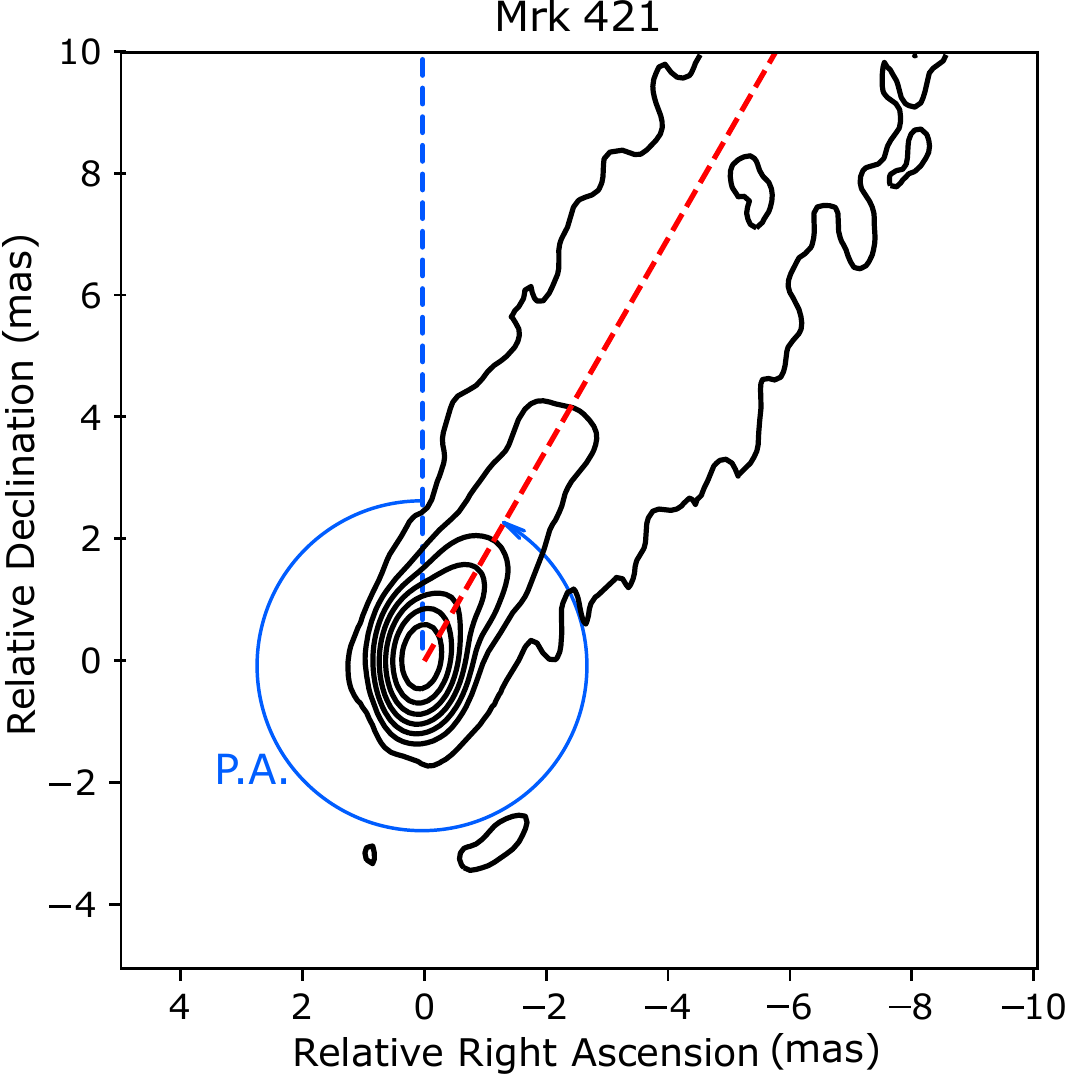}
	\caption{VLBA observation of Mrk 421 at 15 GHz. The red dashed line shows the projected jet direction. The mean position angle ($p.a.$) of the jet feature measured counterclockwise from the right ascension $=0$ line is indicated by the blue angle.}
	\label{Fig:1}
\end{figure}

The MOJAVE program uses the Very Long Baseline Array (VLBA) to monitor northern AGN jets at radio frequencies. To search for the asymmetric signature of offset pair halos, we select AGNs from the MOJAVE sample with well-determined, one-sided jet morphology (Fig. \ref{Fig:1}) with a measured jet position angle \cite{Lister2013,Lister2016}. 

Blazars can be distinguished by their spectral-energy distributions (SEDs). HSP blazars are generally of the BL~Lac type and are characterized by X-ray-peaked synchrotron emission and a hard GeV spectrum. The electron energies needed to produce X-ray synchrotron emission typically imply inverse-Compton emission up to TeV energies. Indeed, TeV emission from these sources is often detected by ground-based $\gamma$-ray telescopes, such as H.E.S.S., MAGIC, and VERITAS \cite{TeVCat}. Therefore, HSP and ISP blazars are more likely to produce TeV $\gamma$-rays that will initiate the pair cascades leading to GeV halos than low-synchrotron-peaked BL~Lacs and quasars. Moreover, at a given cascade energy, a pair halo depends not only on the IGMF strength but also on redshift of the selected source. Thus, we only include HSP and ISP blazars with known redshift in our study.

Comparing the mean free path of pair-production for a TeV photon propagating in the EBL to the light-travel distance from a selected source to the observer, there is a minimum detected energy of the cascade emission, below which the mean free path of pair-production is longer than the source distance. Based on our selection criteria, the lowest redshift of our source candidates, $z \sim 0.03$ (Mrk 421 and Mrk 501), motivates us to look at the Fermi-LAT data at $> 30$ GeV energy (see details in Appendix A). Hence, we further select our sources to show evidence for emission at GeV energies in Fermi-LAT data by requiring association with a source in the 2FHL \cite{2FHL} catalog. To reduce contamination from nearby sources, we only select isolated sources by requiring no nearby 2FHL source with energy $>50$ GeV within $1^\circ$.

Twelve blazars are selected based on the described criteria (Table \ref{Tab:1}). The mean position angle ($p.a.$) of the jet ($\overline{p.a.}$, measured counterclockwise from relative right ascension $=0$ line (pointing in the positive declination direction), and its standard deviation ($\sigma_{p.a.}$) are calculated by averaging the position angle measurements tabulated in \cite{Lister2013,Lister2016}, weighted by observation epochs. 

\section{\label{sec:data}Data preparation and stacked source maps}
To search for GeV halos, $\gamma$-ray events in the Fermi-LAT data (Pass 8 Release 2 Version 6) from mission week 9 (early August, 2008) to 447 (late December, 2016) are selected. We then chose SOURCE-class $\gamma$-ray events with all event types (including both front- and back-converted $\gamma$-ray events) and prepared the data following recommendations from the Fermi Science Support Center \cite{FermiSSC}. The selected data are binned into two energy bins: $30-100$ GeV and $100-300$ GeV.

\begin{figure}[tbp]
	\centering
	\includegraphics[width=8cm, angle=0]{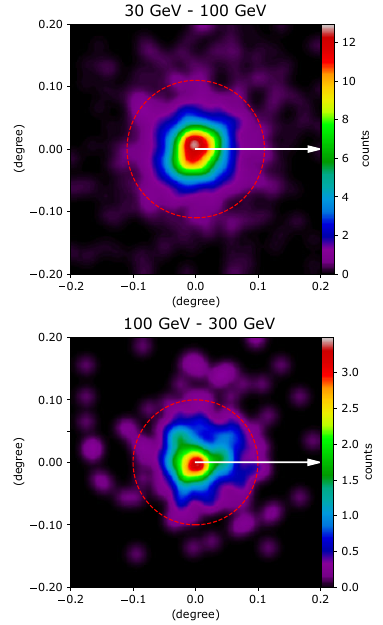}
	\caption{Stacked count maps of $\gamma$-rays detected by the Fermi-LAT. The count maps are calculated with $0.01^\circ$ pixel size and smoothed by convolving with a Gaussian kernel with $\sigma = 0.01^\circ$. White arrows show the position angle of the jets. Red dashed circles are the 68\% containment regions of the Fermi-LAT PSF at the lower bound of each energy range.}
	\label{Fig:2}
\end{figure}

We rotate the sky maps of the Fermi-LAT $\gamma$-ray events to align their jet-$p.a.$ vectors. To visualize the possible asymmetry of the offset pair halos, we stack their Fermi-LAT sky maps after the rotation and convolve the resulting map using a Gaussian kernel with $\sigma = 0.01^{\circ}$ which is much smaller than the 68\% containment of the Fermi-LAT PSF ($\sim 0.1^{\circ}$, \cite{Fermi_performance}). Fig. \ref{Fig:2} shows the resulting sky map in the two energy ranges. The $\gamma$-ray event distribution in the $100-300$ GeV energy range appears to be slightly asymmetric, extending along the jet direction.

\begin{figure}[tbp]
	\centering
	\includegraphics[width=8.5cm, angle=0]{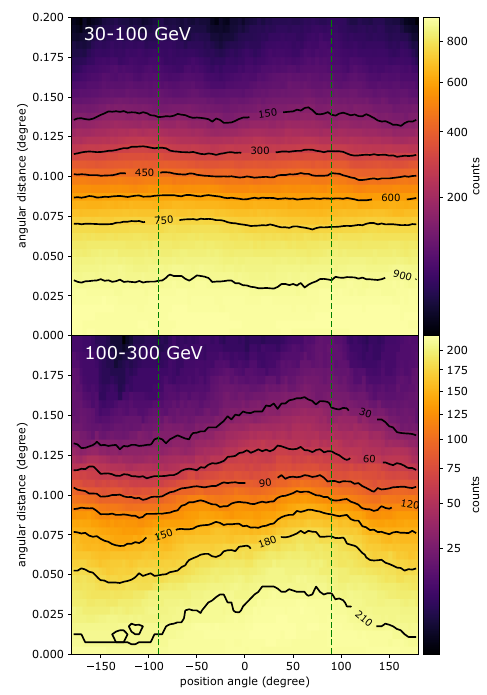}
	\caption{Distribution of summed number of $\gamma$-ray events within a sky patch with $0.1^\circ$ radius around a given position angle from the jet direction and a given angular distance from the source position. Green dashed lines show $\pm 90^\circ$ position angles with respect to the stacked jet (which has position angle $0^\circ$).}
	\label{Fig:3}
\end{figure}

To quantify this apparent asymmetry, we calculate the number of $\gamma$-ray events within a $0.1^{\circ}$ radius region of interest (ROI) at a given angular distance from the source as a function of position angle (Fig. \ref{Fig:3}). Flat contours in Fig. \ref{Fig:3} indicate an azimuthally-symmetric event distribution around the $\gamma$-ray source, as is the case at $30-100$ GeV energies. At higher energies ($100-300$ GeV) the contours peak at $p.a.$ within $90^\circ$ of the projected jet direction (the region between two green-dashed lines in Fig. \ref{Fig:3}), indicating that there are more $\gamma$-rays detected on the jet-side of the sky around the selected blazars than that on the opposite side.

\section{\label{sec:model and test}Pair-halo model and statistical analysis}

\subsection{\label{sec:model}Pair-halo model}

We develop a three-dimensional pair-halo model and perform a likelihood-ratio test (LRT) to determine the statistical significance of the observed asymmetry in the spatial distribution of GeV $\gamma$-rays around the selected blazars. The pair halo model is based on previous studies \cite{Arlen2014,Neronov2010b} but differs in some details of the physics of the pair-halo cascade and approximations made to speed up the computation.

The spatial distribution and intensity of secondary $\gamma$-rays produced in pair halos is a function of the jet emission profile (i.e., the distribution of directions of primary TeV $\gamma$-rays in the jet), jet inclination angle with respect to the line of sight, the redshift of the source, the characteristics of the IGMF, and the spectrum of the intervening EBL and CMB photon fields. We model the jet profile as a two-dimensional Gaussian function with full width at half maximum (FWHM) of $1^\circ$. The choice of angular size is based on a recent study that measured a median intrinsic opening angle of the flows of 135 AGN jets of $1.3^\circ$, with Fermi-LAT-detected AGN having narrower intrinsic opening angle than those that are not detected by Fermi-LAT \cite{Pushkarev2017}. The inclination angle of jet axes with respect to the line of sight (distinct from the jet $p.a.$) is set as a free parameter for each AGN in our sample. Parameters describing the IGMF such as field strength, power spectrum, and helicity remain largely unconstrained. To simplify the simulation, we assume a stochastic non-helical IGMF with coherence length of $\sim 1$ Mpc, which is smaller than the pair-production mean free path of TeV $\gamma$-rays ($\gtrsim 100$ Mpc, derived from Eq. \ref{pp_meanfreepath3} in Appendix A) in the intergalactic medium and larger than the inverse-Compton mean free path of the cascade pair-cooling on the CMB ($\sim 10-100$ kpc, see, e.g., \cite{Neronov2009,Chen2015a}). Hence, the pair cascade process for each primary $\gamma$-ray can be simulated in a uniform magnetic field with random direction and field strength that is set as a free parameter in the model.

\begin{figure}[tbp]
	\centering
	\includegraphics[width=8cm, angle=0]{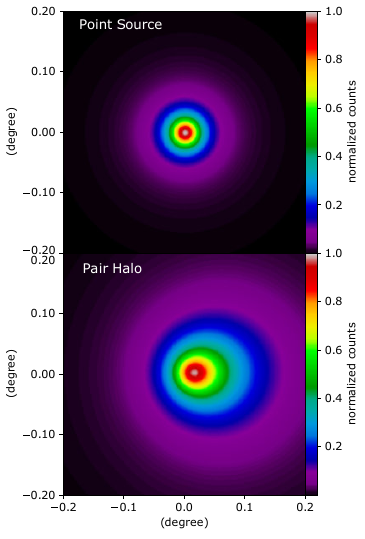}
	\caption{\textit{Top}: Simulated sky maps (normalized count rate convolved with the PSF) of a point source. \textit{Bottom}: the pair halo from a source at $z=0.03$ with jet inclination angle of $1^\circ$ and IGMF strength of $10^{-15}$ G at 100 GeV.}
	\label{Fig:4}
\end{figure}

To calculate the expected number and distribution of detected $\gamma$-ray counts from a pair-halo model, we simulate a large number of detectable cascade $\gamma$-ray events for a given set of jet parameters (jet inclination angle, distance, and position angle) and IGMF strength. Fig. \ref{Fig:4} shows an example of the model result (expectation of normalized counts) of a pair halo at 100 GeV from a source at $z=0.03$ with jet inclination angle of $1^\circ$ and IGMF strength of $10^{-15}$ G.  The point source expectation is given by the Fermi-LAT PSF and the halo expectation is calculated by convolving the simulated arrival photon distribution with the PSF. The model is described in detail in Appendix A.

\subsection{\label{sec:lrt}Likelihood-ratio test for the pair halo hypothesis}

Based on the pair-halo model, we test the pair-halo hypothesis ($H_1$) against the null hypothesis ($H_0$, i.e., an unresolved point source that is described by the LAT PSF) by evaluating their likelihood ratio. We assume the expectation of detected $\gamma$-ray events around a blazar is a linear combination of a point source, a pair halo, and background. Hence, the pair-halo hypothesis is defined in the parameter space of $\boldsymbol{x} =\left(B_0, \{\theta_i\}, \{f_{ij}\}, \{A_{ij}\}, \{\boldsymbol{\mu}_{ij}\}\right)$, where $B_0$ is the present-day IGMF strength, $\theta_i$ is the jet inclination angle for the $i$-th source, $f_{ij}$, $A_{ij}$, and $\boldsymbol{\mu}_{ij}$ are the fraction of the halo component, total source intensity, and background intensity for the $i$-th source in the $j$-th energy bin, respectively. Clearly, the null hypothesis is a special case of the pair-halo hypothesis for either $f_{i,j}=0$ or $B_0=0$. The likelihood ratio can be written as
\begin{equation}
\Lambda(\boldsymbol{x}|\boldsymbol{D})=\\ \frac{\sup\{\mathcal{L}(\boldsymbol{x}, H_1|\boldsymbol{D})\}}{\sup\{\mathcal{L}(\boldsymbol{x}, H_0|\boldsymbol{D})\}},
\label{Eq:1}
\end{equation}
where $\boldsymbol{D}$ denotes the data set and ``$\sup$" is the supremum function. The joint likelihood function for all sources, energies, and pixels is given by
\begin{equation}
\begin{aligned}
\mathcal{L} & \ (\boldsymbol{x}|\boldsymbol{D})= P\Big(\sum^{\mathrm{ROI}}N_{\mathrm{off},ijk}|\sum^{\mathrm{ROI}}\mu_{ijk}(\boldsymbol{x})\Big) \\ & \ \times\prod_i\prod_j\prod_k P\Big(N_{\mathrm{on},ijk}|\lambda_{ijk}(\boldsymbol{x})\Big),
\end{aligned}
\label{Eq:2}
\end{equation}
where $P(N|\lambda)$ denotes a Poisson distribution for getting $N$ counts with expectation $\lambda$. $N_{\mathrm{on},ijk}$ and $N_{\mathrm{off},ijk}$ are the number of source and background events in the $j$-th energy bin and the $k$-th pixel around the $i$-th source. $\lambda_{ijk}$ and $\mu_{ijk}$ are expectations of $N_{\mathrm{on},ijk}$ and $N_{\mathrm{off},ijk}$, respectively, which can be given by the pair-halo model for a given set of parameters, $\boldsymbol{x}$. It is extremely computational intensive to model the pair halo for a large number of $B_0$-$\{\theta_i\}$ combinations at many different energies. To reduce the computational time for this test, we model the pair halo using the lower bound of each energy bin, i.e., we only apply our model at 30 and 100 GeV energies in the LRT. 

\begin{figure}[tbp]
	\centering
	\includegraphics[width=8cm, angle=0]{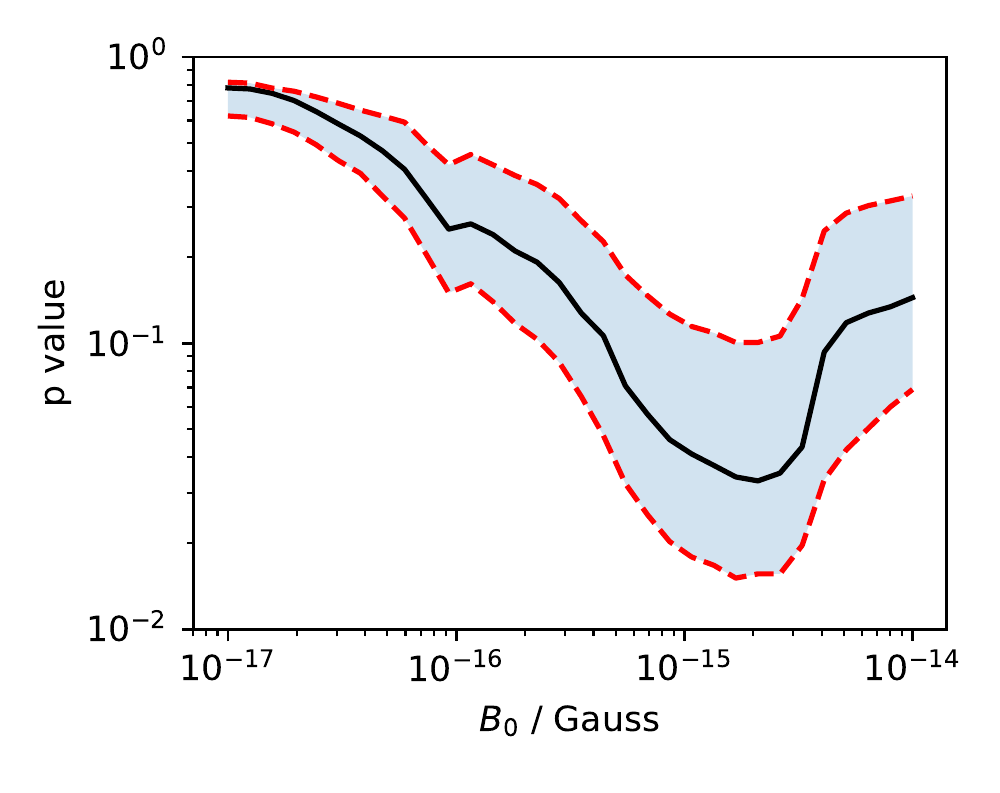}
	\caption{$p$-value as a function of the present-day IGMF strength estimated using the simulated $TS$ distribution. The solid black line is the result using the average $p.a.$ (i.e., $\overline{p.a.}$). The region in between the two dashed red lines shows the 95\% interval of the resulting $p$-values assuming the $p.a.$ is randomly drawn from a normal distribution with mean and standard deviation given by $\overline{p.a.}$ and $\sigma_{p.a.}$, respectively.}
	\label{Fig:5}
\end{figure}

Considering that the 68\% containment angle of the Fermi-LAT PSF in our selected energy range is $\sim 0.1^\circ$, we bin $\gamma$-rays events in each energy range into a $0.8^\circ \times 0.8^\circ$ sky map with pixel size of $0.02^\circ$ centered at the VLBA position of each source. Binning was determined \emph{a priori} with the goal of obtaining adequate statistics. Since we selected $\gamma$-rays at energies $>30$ GeV from extragalactic sources, the number of background counts within our sub-degree-scale ROI is negligible compared to the events recorded from identified Fermi-LAT sources in the extragalactic sky. This allows us to assume that all the detected $\gamma$-ray events are from the known Fermi-LAT sources. We also assume that the estimators of background counts $\mu_{ijk}$ are isotropic for all the pixels in the ROI, which can be given by a single parameter $\mu_{ij}$ around the $i$-th source in the $j$-th energy bin.

We then evaluate a test statistic $TS\equiv 2\ln\Lambda$ for this LRT of the Fermi-LAT data in the parameter space of $\boldsymbol{x}$, where $\Lambda$ is given by Eq.\ref{Eq:2}. 
$TS$ values cannot be directly interpreted in therms of significance of the tested hypothesis because Wilks's theorem is not valid when the null-hypothesis is located at the boundary of the parameter space that models are allowed to explore (see discussion in 
Appendix B). Instead, we 
interpret the statistical significance of our test statistic 
by simulating pair-halo distributions and 
calculating the $TS$ for each MC sample. Details of the simulation procedure are described in Appendix B.

In this study, we are specifically interested in the dependence of the pair halo on the IGMF rather than a detailed characterization of the parameters of the AGN jet. Therefore we marginalize over all jet parameters to determine the $TS$ as a function of the IGMF strength. Fig. \ref{Fig:5} shows the resulting $p$-values as a function of the IGMF strength. We find a minimum $p$-value of $\sim 0.033$ for a pair-halo model with  IGMF strength of $\sim 10^{-15}$ G, corresponding to a 2-$\sigma$ significance. Confidence intervals are computed by randomizing the jet orientation by $p.a. \sim \mathcal{N}(\overline{p.a.}, \sigma_{p.a.}^2)$, where $\mathcal{N}(a, b^2)$ denotes a normal distribution with mean and standard deviation given by $a$ and $b$, respectively.

\begin{figure}[tbp]
	\centering
	\includegraphics[width=7cm, angle=0]{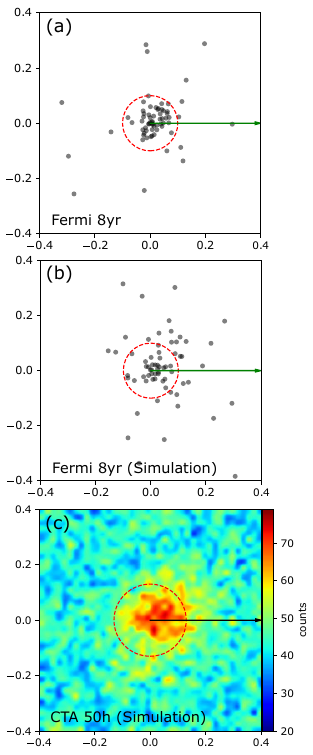}
	\caption{(a) 8-year Fermi-LAT observation (Pass 8 SOURCE events) of Mrk 501 at $100-300$ GeV. (b) A simulation of Mrk 501 8-year count map at 100 GeV observed by the Fermi-LAT assuming a pair halo with $f_\mathrm{halo}=1$ based on the count rate of the 8-year Fermi-LAT observation and assuming an IGMF with strength of $\sim 10^{-15}$ G and a jet with $1^\circ$ inclination angle with respect to the line of sight. (c) A simulation of Mrk 501 50-hour count map (calculated with $0.01^\circ$ pixel size) at 100 GeV that might be observed by the future CTA ground-based $\gamma$-ray observatory using the same halo, IGMF and jet parameters as used in (b) and assuming the effective area of CTA is $8\times 10^4 \mathrm{m}^2$ and the 68\% containment angle of the CTA PSF is $0.13^\circ$. Arrows show the jet orientation given by VLBA observations. Red-dashed circles are the 68\% containment angles of the Fermi-LAT and CTA PSFs at 100 GeV. $x$- and $y$-axes are in units of degree.}
	\label{Fig:6}
\end{figure}

\section{\label{sec:discussion}Discussion}

The joint-likelihood analysis of our sample of 12 HSP/ISP blazars shows no strong evidence extended pair-halo emission at GeV energies. A hint of asymmetric excess can be seen in the $30-300$ GeV energy range with a global $p$-value equivalent to a 2-$\sigma$ significance for getting a normal distributed variable. The maximum likelihood is given by a pair-halo model with IGMF strength of $\sim 10^{-15}$ G. This value is consistent with the inferred range of IGMF strength from a prior study that used stacked HBLs at energies $\sim 1$ GeV \cite{Chen2015a}, and also compatible with estimates and constraints from other independent studies \cite{Neronov2010a,Neronov2011,Essey2011,Tanaka2014}.

A number of assumptions, simplifications, and free parameters are embedded in the pair-halo model. Some are largely underconstrained, such as those parameterizing the IGMF, TeV jet geometry, and the fraction of the cascade emission in the total detected $\gamma$-rays. More complex IGMF configurations than the one described in Section \ref{sec:model}, such as a helical field, may alter the expected pair-halo distribution \cite{Long2015,Batista2016,Duplessis2017}. Moreover, we are not able to estimate the fraction of the cascade emission in the detected $\gamma$-ray beam (parameterized as $f_\mathrm{halo}$) because the intrinsic GeV-TeV spectrum of BL Lacs is uncertain, and it is unclear whether processes other than inverse-Compton contribute to the cooling of the pair cascade \cite{Broderick2012,Schlickeiser2012,Miniati2013,Sironi2014,Kempf2016}. Hence, $f_\mathrm{halo}$ is set as a free parameter in each energy bin for each source in the statistical test.

\begin{figure}[tbp]
	\centering
	\includegraphics[width=8cm, angle=0]{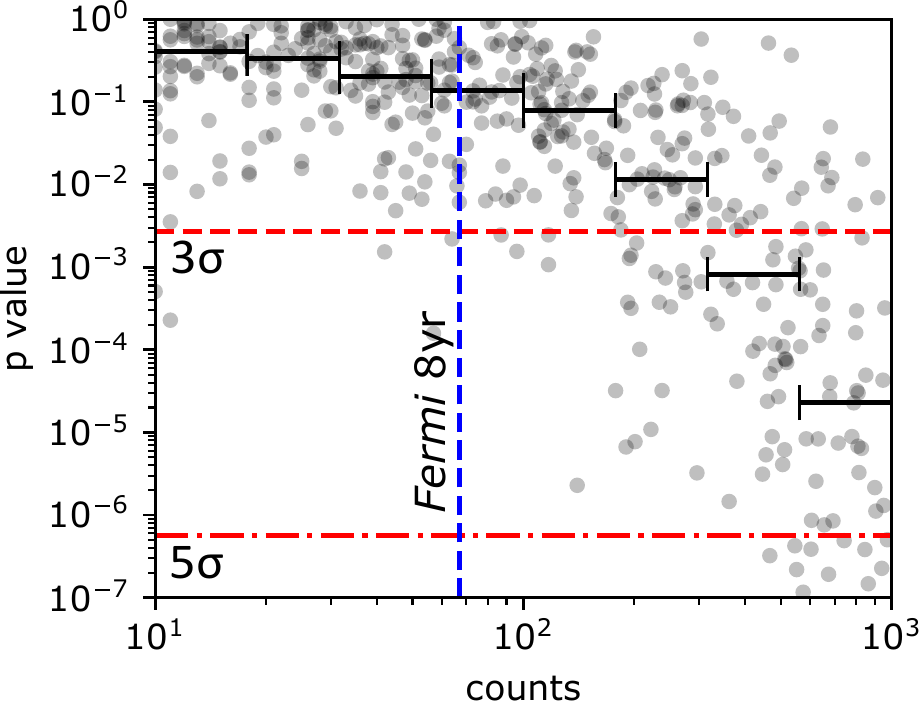}
	\caption{$p$-values (scatters) given by testing a set of simulated Mrk 501 at 100 GeV from Fermi-LAT with (randomly generated) increasing count rate assuming a pair halo with $f_\mathrm{halo}=0.2$, an IGMF with strength of $10^{-15}$ G and a jet with $1^\circ$ inclination angle with respect to the line of sight. Black horizontal bars show median of the $p$-value in each logarithm $x$-bin. Dashed blue line shows the Mrk 501 count rate of the 8-year Fermi-LAT observation in the $100-300$ GeV energy range. Dashed and dash-dotted red lines are the $3\sigma$ and $5\sigma$ confidence level (in terms of $p$-value) for a normal distributed variable, respectively.}
	\label{Fig:7}
\end{figure}

An additional limitation of our analysis is the uncertainty in the geometry of TeV jets. Angular profiles different from a two-dimensional Gaussian would affect the resulting pair-halo. In addition, there is a considerable uncertainty in position-angle measurements of radio jets (see $\sigma_{p.a.}$ values in Table \ref{Tab:1}). If the pair-halo signal is physical, the significance of the detection could be improved with a more precise knowledge of the jet position angle. Furthermore, VLBA observations map the geometry of the jet at kiloparsec scales, while TeV emission is generally thought to originate closer to the central supermassive black hole \cite{Marscher2008}. Differences between the direction of the jet axis in the region where the TeV emission originates and the large-scale jet mapped by radio observations would dilute the significance of a potential halo signal. Systematic surveys at higher radio frequencies and a better understanding of the location and geometry of TeV-emitting region in AGN jets would reduce the systematic uncertainty of future pair-halo searches.

For typical jet parameters, IGMF values and source redshifts, a pair-halo signal is best resolved at energies $\sim 100$ GeV. The effective area of Fermi-LAT results in a small number of counts at these energies, but future space missions and ground-based observatories such as the Cherenkov Telescope Array (CTA) will certainly produce larger event statistics. To illustrate the potential of future ground-based experiments in pair-halo searches, we simulate datasets using our model for Mrk~501, one of the brightest sources in our sample. As shown in Fig. \ref{Fig:6}, count maps of the pair halo of Mrk~501 ($f_\mathrm{halo}=1$) at $100-300$ GeV energy from Fermi-LAT and CTA are simulated assuming an IGMF with strength $10^{-15}$ G and a jet with $1^\circ$ inclination angle with respect to the line of sight. Since the pair halo spectrum is unknown, to simplify the simulation we derive the count rate from the 8-year Fermi-LAT observations in the $100-300$ GeV energy range but assume all simulated pair-halo $\gamma$-rays are at 100 GeV energy. So, we apply our pair-halo model and use Fermi-LAT and rough CTA performance values only at 100 GeV. For the CTA simulation, we assume a 50-h exposure with a background rate of $0.6\ \mathrm{Hz}/\mathrm{degree}^2$, effective area of $8\times 10^4 \mathrm{m}^2$, and a PSF described by a two-dimensional Gaussian function with 68\% containment angle of $0.13^\circ$ \cite{CTA_performance}. We assume the extreme case that $f_\mathrm{halo}=1$. As we can see, CTA observations provide a much larger sample size of $\gamma$-ray events at VHE energies and could yield better measurements of an offset halo if systematic errors in the PSF can be controlled.  

To test how much more data it would take to confirm the hint of halo signal we found in our study with a future space-based instrument, we simulate a number of Mrk~501-like sources with increasing number of counts at 100 GeV energy from Fermi-LAT assuming $f_\mathrm{halo}=0.2$ and using the same IGMF and jet model parameters. We then perform our LRT test for each simulated dataset and study the evolution of the $p$-value as event statistics increase (Fig. \ref{Fig:7}) \cite{note2}. Based on results in Fig. \ref{Fig:7}, we expect that a mission capable of collecting $\gtrsim 10$ times the 8-year Fermi-LAT exposure at 100 GeV would be able to detect the extended halo produced by a $10^{-15}$ G IGMF at 5$\sigma$ level based on the analysis of a single source.

While hints of the presence of a pair halos continue to accumulate in the literature, a positive indirect detection of the IGMF continues to be elusive due to our limited knowledge of the astrophysics of TeV-emitting blazars and the fact that the expected angular size of pair halos is comparable to the angular resolution of current gamma-ray observatories. A better understanding of the physical processes that give rise to TeV emission in relativistic jets, together with a new generation of instruments with improved angular resolution,  may finally yield a positive detection of pair halos in the coming years.

\begin{acknowledgments}
	This work was partially supported by NASA grant NNH14ZDA001N. Authors acknowledge the Fermi team for providing the Fermi-LAT data. This research has made use of the CTA instrument response functions provided by the CTA Consortium and Observatory. This research has made use of data from the MOJAVE database that is maintained by the MOJAVE team.
\end{acknowledgments}

\appendix
\section{Simulation of Pair Halos}

\subsection{Physics Processes}

AGN are the most numerous class of extragalactic sources in the high-energy (HE, GeV) and very-high-energy (VHE, $>100$ GeV) bands. $\gamma$-rays  are thought to be generated via the IC scattering process, by which the relativistic particles (electrons and positrons) up-scatter the ambient infrared and optical photons or the UV--X-ray synchrotron photons up to $\gamma$-ray energies. As VHE $\gamma$-rays travel through intergalactic space, pair production between TeV $\gamma$-rays and diffuse photons from the extragalactic background light (EBL) attenuates the VHE photon flux while creating relativistic electron-positron pairs. The survival probability $P(l)$ for a TeV photon to propagate a distance $l$ through the intergalactic medium is given by
\begin{equation}
P(l)=P(l_0)\exp\left(-\int_{l_0}^{l}\frac{\mathrm{d}l}{D_{\gamma}}\right),
\label{MC_pp_length}
\end{equation}
where $D_{\gamma}$ is the mean free path for pair production. $D_{\gamma}$ is a function of the incident $\gamma$-ray photon's energy $E_{\gamma_0}$, redshift $z$, the cross section of $\gamma\gamma$ pair production $\sigma_{\gamma\gamma}$ and the number density of EBL photons $n_\mathrm{EBL}(\epsilon_\mathrm{EBL},z)$, where $\epsilon_\mathrm{EBL}$ is the EBL photon energy \cite{Franceschini2008,FermiEBL2012}). Following the discussion in Ref \cite{Neronov2009}, a significant uncertainty appears in estimating $D_{\gamma}$ by adopting different EBL models, but they all agree that $D_{\gamma}$ decreases with the increasing $E_{\gamma_0}$ and $z$. We use the simplified expression for $D_\gamma$ from \cite{Neronov2009}:
\begin{equation}
D_\gamma\approx 80\frac{\kappa}{(1+z)^2}\left(\frac{E_{\gamma_0}}{10\mathrm{TeV}}\right)^{-1}\mathrm{Mpc},
\label{pp_meanfreepath3}
\end{equation}
where $\kappa\sim 1$ accounts for the EBL model uncertainties. The energy of the resulting electron/positron in the laboratory frame (rest frame of observer) can be taken to be approximately the kinematic maximum value
\begin{equation}
E_e = \frac{E_{\gamma_0}+\epsilon_\mathrm{EBL}}{2} \approx \frac{E_{\gamma_0}}{2},
\label{E_e}
\end{equation}
since $E_{\gamma_0} \gg \epsilon_\mathrm{EBL}$.

While there is some debate over the role of plasma processes in energy loss \cite{Broderick2012,Schlickeiser2012,Miniati2013,Sironi2014,Kempf2016}, we assume that the pairs cool through IC scattering with CMB and EBL photons. As a result, the background photons are upscattered into the $\gamma$-ray energy range. In a collision between a relativistic electron and a CMB photon, the center-of-mass frame is very close to the rest frame of the electron. The energy of the CMB photon in this frame is given by
\begin{equation}
\hbar \omega' = \gamma\hbar\omega\left[ 1+\frac{v}{c}\cos\theta\right] \sim \gamma\hbar\omega,
\label{ThomsonCondition1}
\end{equation}
where $\gamma$ is the Lorentz factor, $\theta$ is the angle between the directions of the two particles in the laboratory frame, $\omega'$ and $\omega$ are the angular frequency of the CMB photon in the rest frame of the electron and in the laboratory frame, respectively. We are interested in comparing this energy with the electron energy in the center-of-mass frame, which is very close to its rest energy $m_ec^2$. For $E_{\gamma_0} \approx 10$ TeV we have
\begin{equation}
E_e\epsilon_{\mathrm{CMB}}\approx3\times10^9\mathrm{eV}^2\ll(m_ec^2)^2\approx2.5\times10^{11}\mathrm{eV}^2,
\label{ThomsonCondition2}
\end{equation}
where $\epsilon_{\mathrm{CMB}}$ is the typical energy of a CMB photon. As $\gamma\hbar\omega \ll m_ec^2$ the interaction happens in the Thomson regime, resulting in an energy loss rate of the pair beam is given by
\begin{equation}
-\frac{\mathrm{d}E}{\mathrm{d}t}=\frac{4}{3}\sigma_\mathrm{T}cU_\mathrm{rad}\left(\frac{v^2}{c^2}\right)\gamma^2,
\label{energylossrate_pair}
\end{equation}
where $U_\mathrm{rad}$ is the energy density of radiation in the laboratory frame and $\sigma_\mathrm{T}$ is the Thomson scattering cross section. Since the energy density of the CMB is much higher than the density of the infrared/optical background (EBL), $U_\mathrm{rad}$ is dominated by the CMB energy density $U_\mathrm{CMB}'$ at the redshift $z_{\gamma\gamma}$ where the IC occurs. For relativistic electrons with $v \approx c$, we have
\begin{equation}
U_\mathrm{CMB}' = \epsilon_{\mathrm{CMB}}'n_{\mathrm{CMB}}' = \frac{\epsilon_{\mathrm{CMB}}'\mathrm{d}N_\mathrm{CMB}}{\mathrm{d}V}=\frac{\epsilon_{\mathrm{CMB}}'\mathrm{d}N_\mathrm{CMB}}{\sigma_\mathrm{T}c\mathrm{d}t},
\label{Ucmb}
\end{equation}
where $\epsilon_{\mathrm{CMB}}'$ and $n_{\mathrm{CMB}}'=\mathrm{d}N_\mathrm{CMB}/\mathrm{d}V$ are the typical energy and number density of CMB photons at redshift $z_{\gamma\gamma}$, respectively. Hence, the energy of cascade $\gamma$-rays produced by IC scattering, as observed on Earth, is given by
\begin{equation}
\begin{aligned}
E_\gamma = & \ -\frac{\mathrm{d}E}{\mathrm{d}N_\mathrm{CMB}} (1+z_{\gamma\gamma})^{-1} \\ \approx & \ \frac{4}{3}(1+z_{\gamma\gamma})^{-1}\epsilon_{\mathrm{CMB}}'\left(\frac{E_e}{m_ec^2}\right)^2.
\end{aligned}
\label{energyofcascade1}
\end{equation}
Inserting $\epsilon_{\mathrm{CMB}}'=6\times 10^{-4}(1+z_{\gamma\gamma})\mathrm{eV}$, we obtain
\begin{equation}
E_\gamma\approx 77\mathrm{GeV}\left(\frac{E_{\gamma_0}}{10\mathrm{TeV}}\right)^2.
\label{energyofcascade2}
\end{equation}
The mean free path for producing one cascade $\gamma$-ray via IC scattering at redshift $z_{\gamma\gamma}$ is give by
\begin{equation}
D_\mathrm{IC}=\left< \sigma_\mathrm{T} n_\mathrm{CMB}' \right>^{-1}.
\label{meanfreepathIC0}
\end{equation}
Assuming the total number of CMB photons is conserved so that the evolution of the CMB number density follows $n_\mathrm{CMB} \propto a^{-3} \propto (1+z)^3$, where $a$ is the scale factor. Plugging the present-day CMB number density, $\sim 413\ \mathrm{cm}^{-3}$, into Eq.(\ref{meanfreepathIC0}), we have
\begin{equation}
D_\mathrm{IC} \approx 3.6\times 10^{19} (1+z_{\gamma\gamma})^{-3}\ \mathrm{m} \approx 1.2\ (1+z_{\gamma\gamma})^{-3}\ \mathrm{kpc}.
\label{meanfreepathIC}
\end{equation}

\begin{figure}[tbp]
	\centering
	\includegraphics[width=8.5cm, angle=0]{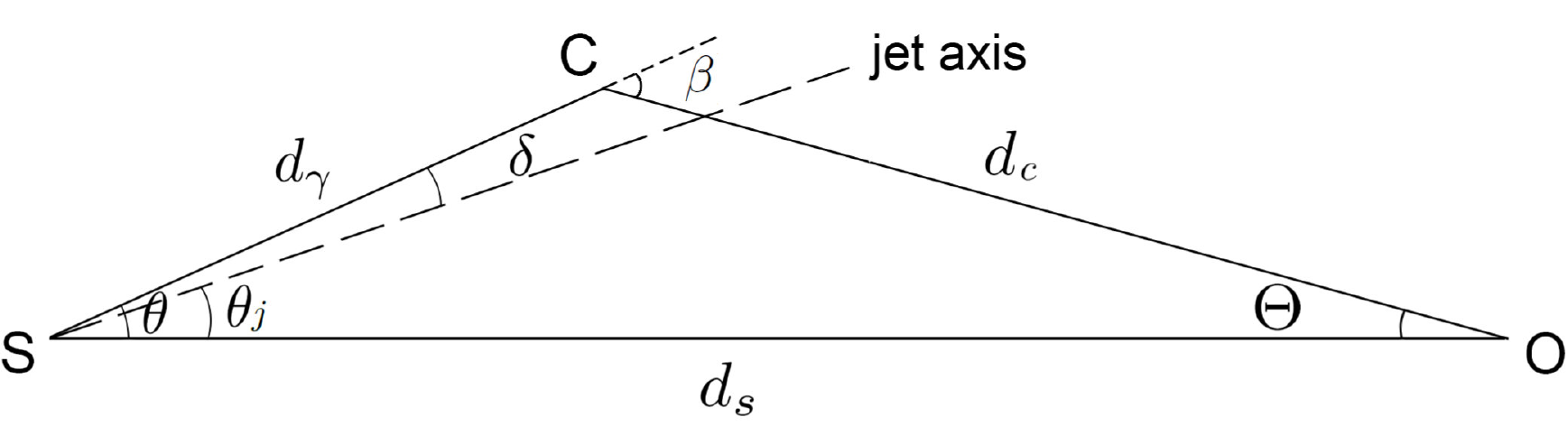}
	\caption{\label{Fig:S1}Geometry of the pair-halo emission showing propagation of direct and cascade $\gamma$-rays. The pair trajectory is not shown since this distance is much smaller than the $\gamma\gamma$ pair creation distance.}
\end{figure}

\begin{figure}[b]
	\centering
	\includegraphics[width=8cm, angle=0]{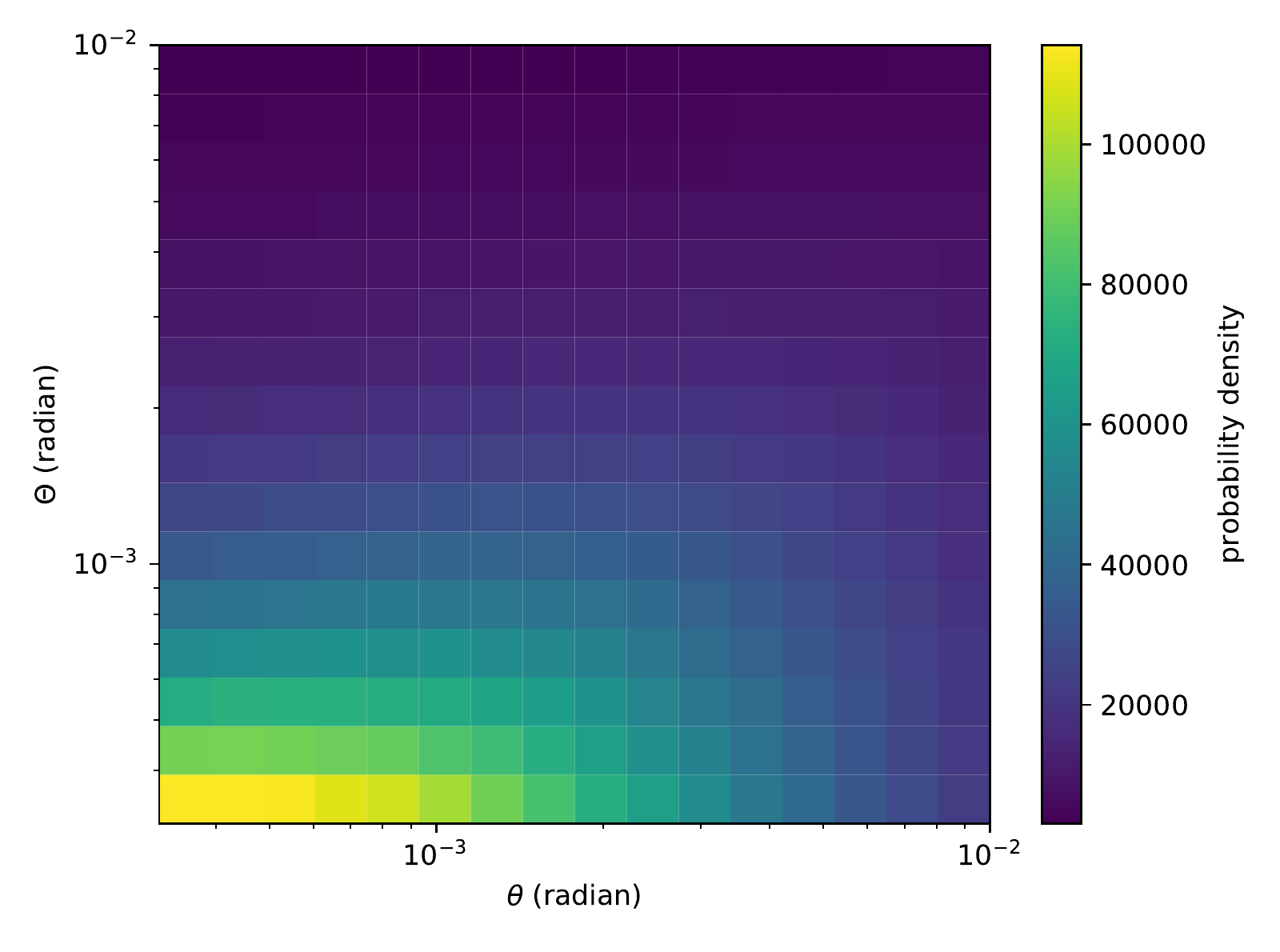}
	\caption{\label{Fig:S2}Simulated probability density distribution as a function of $\theta$ and $\Theta$ (angles as illustrated in Fig. \ref{Fig:S1}) for $E_\gamma=10$ GeV, $z=0.1$, $B_0=10^{-15}$ G, and $\theta_j=3^\circ$.}
\end{figure}

Let us also define a half-cooling length, $D_e$, as the propagating distance in which a relativistic electron loses half of its energy $E_e/2$ via multiple IC scattering interactions. By multiplying $v^{-1}=\mathrm{d}t/\mathrm{d}x$ on both sides of Eq.(\ref{energylossrate_pair}), the equation can be rewritten as
\begin{equation}
-\frac{\mathrm{d}E}{\mathrm{d}x} = \frac{4}{3}\sigma_\mathrm{T}U_\mathrm{CMB}'\gamma\sqrt{\gamma^2-1}.
\label{energylossrate_pair0}
\end{equation}
The Lorentz factor $\gamma$ is a function of $E$, and hence Eq.(\ref{energylossrate_pair0}) can be rearranged as
\begin{equation}
-(E^4-E^2m_e^2c^4)^{-1/2}\mathrm{d}E = \frac{4\sigma_\mathrm{T}U_\mathrm{CMB}'}{3m_e^2c^4}\mathrm{d}x.
\label{energylossrate_pair1}
\end{equation}
Integrating both sides over the energy range from $E_e$ to $E_e/2$, and noticing that $E_e \gg m_ec^2$ for VHE electrons, $D_e$ can be given by

\begin{equation}
\begin{aligned}
D_e \approx & \ \frac{3m_e^2c^4}{4\sigma_\mathrm{T}U_\mathrm{CMB}'E_e} \\ \approx & \ 1.2\times 10^{21}(1+z_{\gamma\gamma})^{-4}\left(\frac{E_e}{10\mathrm{TeV}}\right)^{-1}\mathrm{m} \\ \approx & \ 2.4\times 10^{21}(1+z_{\gamma\gamma})^{-4}\left(\frac{E_{\gamma_0}}{10\mathrm{TeV}}\right)^{-1}\mathrm{m} \\ \approx & \ 77.8\ (1+z_{\gamma\gamma})^{-4}\left(\frac{E_{\gamma_0}}{10\mathrm{TeV}}\right)^{-1} \mathrm{kpc}.
\end{aligned}
\label{D_e}
\end{equation}
If the cascade $\gamma$-rays are energetic enough, they will keep pair producing and creating the next generation $\gamma$-rays via IC scattering until they are cooled to sub-TeV energies. The mean free path of pair production for the next generation cascade $\gamma$-ray is much longer than that for the previous generation because of the significantly lower energies. At $\lesssim 100$ GeV energy, the mean free path of pair production is so long (in the scale of the Hubble radius) that no additional cascades will be likely to happen in our detecting range.

\subsection{Pair-Halo Model}

In intergalactic space, magnetic fields deflect the electron-positron pairs, changing the angular distribution of the secondary cascade emission. The geometry of the cascade interaction is shown in 
Fig. \ref{Fig:S1}. Assuming a blazar located at point $S$, we (the observer) are at point $O$ with a comoving distance $d_s$ from the source. 
An off-axis TeV $\gamma$-ray emitted at $S$ produces an electron-positron pair at point $C$ and the IGMF bends one charged particle 
back toward our line of sight. We thus receive a secondary $\gamma$-ray with an off-axis angle $\Theta$. 
We assume 
$d_s \gg D_\gamma$. 
Typically, $D_e$ (from Eq. (\ref{D_e})) is much smaller than $d_\gamma$ and $d_s$.

\begin{figure}[tbp]
	\centering
	\includegraphics[width=8cm, angle=0]{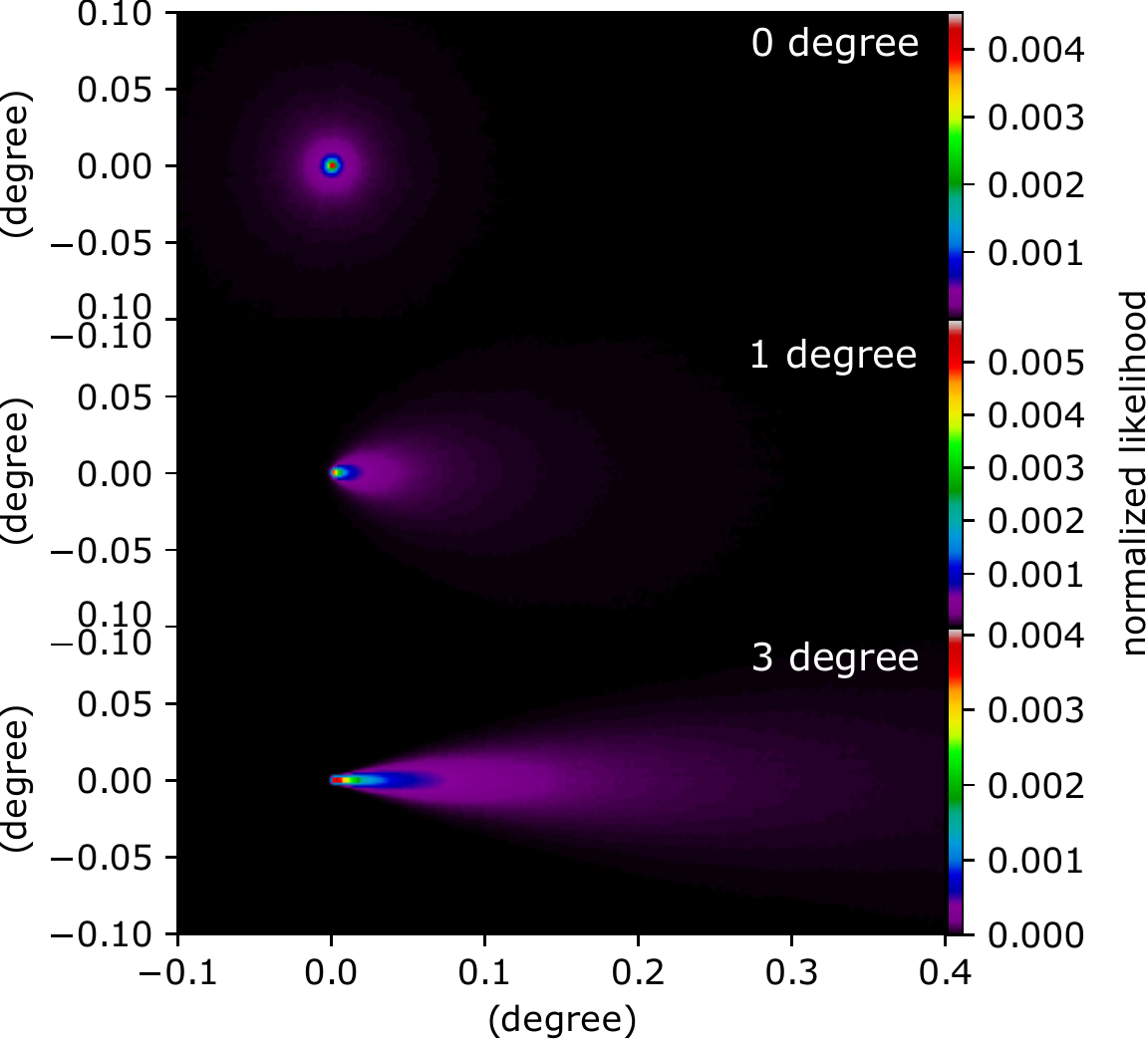}
	\caption{\label{Fig:S3}Likelihood of the pair-halo count rate at 10 GeV simulated using $10^6$ primary TeV $\gamma$-rays from a source at $z=0.1$ with IGMF strength of $10^{-15}$ G and jet inclination angle of $0^\circ$, $1^\circ$, and $3^\circ$, respectively. Offset jet-axes are orientated to the right.}
\end{figure}

The IGMF is described as a stochastic non-helical magnetic field with coherence length $\ell\sim 1$ Mpc, such that $D_e \ll \ell \ll D_\gamma$.  That is much smaller than the mean free path of the pair production for TeV $\gamma$-rays ($\gtrsim 100$ Mpc, from Eq. \ref{pp_meanfreepath3}) and larger than the typical IC cooling length for the pairs in CMB fields ($\sim 10-100$ kpc, from Eq. \ref{D_e}). 
Hence, 
the pair cascade process for each primary $\gamma$-ray can be simulated in a uniform magnetic field with a random direction. 
Fig. \ref{Fig:S2} shows an example of the simulated probability density distribution $f(\theta, \Theta)$.
We see the probability density is highly peaked along the line of sight where $\theta=0^\circ$ (note that axes of Fig. \ref{Fig:S2} are in logarithmic scale), indicating that pair halos are peaked at point-source positions, although their signals are broadened around the sources.

\begin{figure}[t]
	\centering
	\includegraphics[width=8.5cm, angle=0]{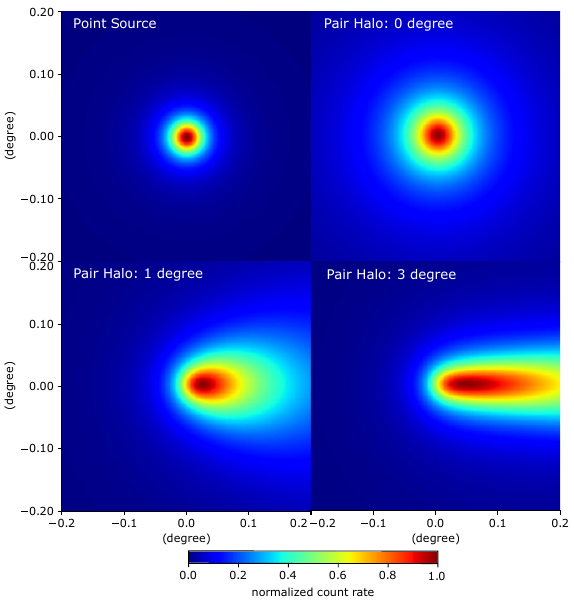}
	\caption{\label{Fig:S4}Expectation of point-source and pair-halo count rate in the presence of the Fermi-LAT PSF at 10 GeV. Pair halos are simulated using $10^6$ primary TeV $\gamma$-rays from a source at $z=0.1$ with IGMF strength of $10^{-15}$ G and jet inclination angle of $0^\circ$, $1^\circ$, and $3^\circ$, respectively. Offset jet-axes are orientated to the right.}
\end{figure}

Primary TeV $\gamma$-rays are 
collimated along the jet axis. We assume the TeV-jet has the similar opening angle as the radio jet, 
and model the angular distribution of primary $\gamma$-rays with a two-dimensional Gaussian with $1^\circ$ full width at half maximum (FWHM). Fig. \ref{Fig:S3} shows an example of the model result of halos at 10 GeV. 
Expected halo images recorded by \emph{Fermi}-LAT 
are shown in Fig. \ref{Fig:S4}.
Our model provides very similar pair-halo morphologies than 
previous studies, e.g., \cite{Neronov2010b,Arlen2014}.

\begin{figure}[t]
	\centering
	\includegraphics[width=7.5cm, angle=0]{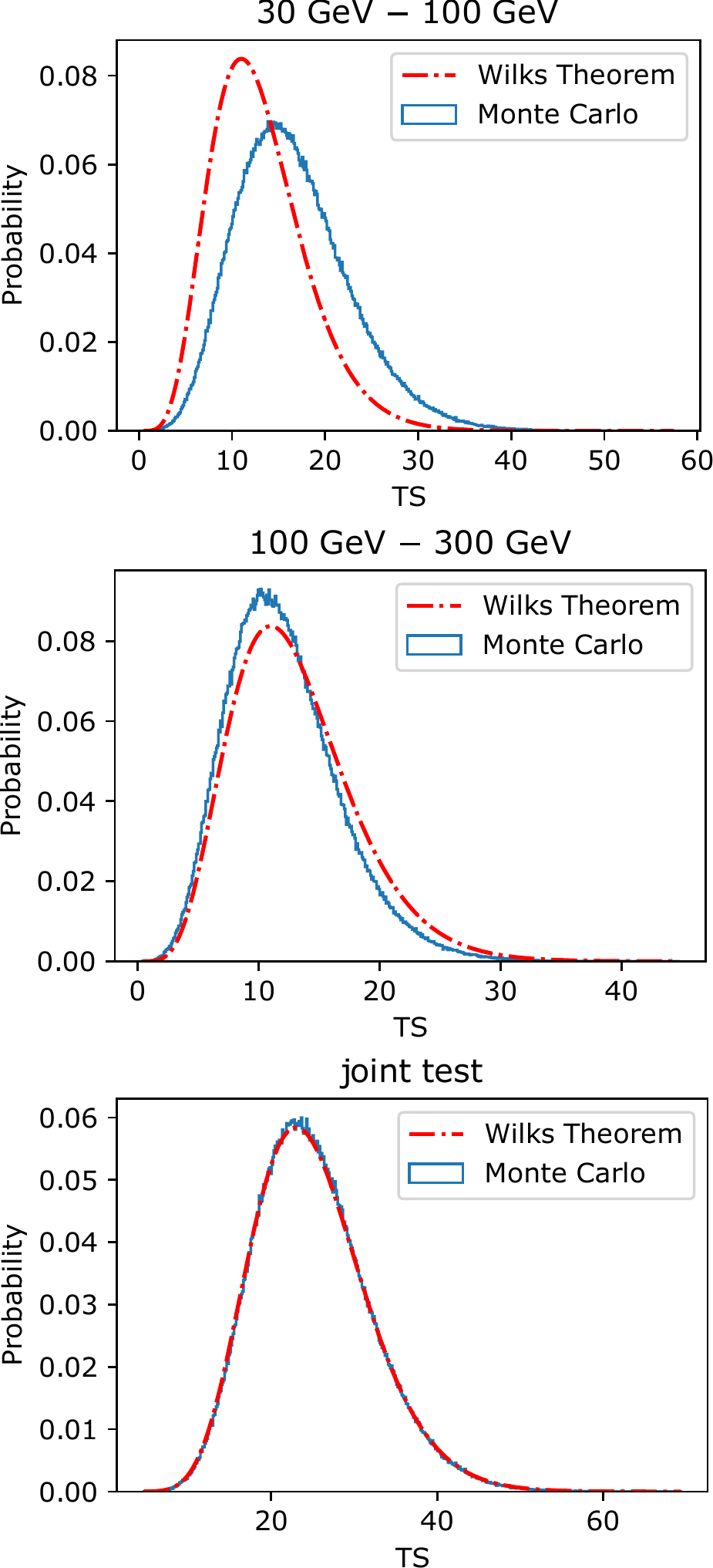}
	\caption{\label{Fig:S5}Probability distribution of the test statistic $TS\equiv 2\ln\Lambda$ simulated based on the null hypothesis $H_0$ using a Monte Carlo method. Red dashed lines are the $\chi^2$-distributions given by Wilks' theorem: $\chi^2_{13}$ for $30-100$ GeV and $100-300$ GeV energy bins, $\chi^2_{25}$ for the joint test.}
\end{figure}

In our likelihood-ratio test (LRT) of offset pair halos, we need to go through a wide range of values in the parameter space. For each source with a known redshift at a given energy, every combination of $B_0$ and $\theta_j$ values corresponds to an independent simulation of a pair halo. The previous examples with $10^6$ samples take $\sim 20$ minutes  for each pair-halo simulation based on a multiple-threaded CPU computation using 8 threads \cite{note3}. If we are going to test a source at a given energy in the parameter space with $32\times 32$ choices of $B_0$ and $\theta_j$, the likelihood-ratio test becomes computational difficult. We note that the pair-halo model consists of a large number of independent cascade-event simulations which is highly parallelizable. In this study, we use the massively parallel processing power of the GPU to boost performance of the pair-halo model. In detail, we use CUDA Toolkit 9.2 \cite{cuda9.2} to parallelize our python code. We subdivide the $10^6$ samples into 100 computing groups and assign each group to a GPU grid. We use $10\times 1$ blocks per grid and $1024\times 1$ threads per block to perform the simulation. The GPU-based code is much faster than our CPU version, which takes $\sim 16$ seconds to simulate each pair halo as shown in the previous examples (Fig. \ref{Fig:S3}, \ref{Fig:S4}).

~

\section{Simulation of the LRT's Test Statistic}

For a LRT of hypothesis $H_1$ against the null hypothesis $H_0$, the likelihood ratio is given by Eq. \ref{Eq:1}.
According to the Wilks theorem, the distribution of a test statistic $TS \equiv 2\ln{\Lambda}$ will be asymptotically $\chi^2$-distributed with the number of degrees of freedom equal to the difference in dimensionality of $H_1$ and $H_0$ as the sample size of the repeatable data sets $\boldsymbol{D}$ goes to infinity. However, the theorem is only valid if $H_1$ is an open set and $H_0$ is a subset of $H_1$ which has at least one more degree of freedom than $H_0$ (see a discussion in, e.g., \cite{Protassov2002}). Unfortunately, the theorem is not valid for this study because the parameter domain of $H_1$ is clearly not open and the null hypothesis $H_0$ associated with the non-detection of pair halos, is located on the boundary of the $H_1$ parameter domain (i.e., either $f_{i,j}=0$ or $B_0=0$, as described in the main text). In order to interpret the statistical significance of our test statistic $TS$, we simulate the distribution of $TS$ by generating a number of sources and calculating $TS$ for each Monte Carlo (MC) sample. The $p$-value is then estimated from the simulated $TS$ distribution using the $TS$ value obtained from the LRT with \emph{Fermi} observations.


A MC sample is a group of simulated sources based on the null hypothesis. In detail, each MC sample consists of a set of simulated sources at a given energy whose $\gamma$-ray counts in each pixel are generated based on Poisson distributions with expected counts following the null hypothesis $H_0$, i.e., the \emph{Fermi} PSF. For the simulated sources, the total expected counts over each source are the same as the averaged count rate given by the observation. We apply the same LRT as we used for the real data to each of the MC samples and the distribution of the $TS$ is then given by testing a large number of MC samples. Fig. \ref{Fig:S5} shows the resulting $TS$ distributions from our MC simulation. Results of the Wilks' theorem, i.e., $\chi^2$-distributions with the number of degrees of freedom equal to the difference in dimensionality of $H_1$ and $H_0$, are also shown. As we can see from Fig. \ref{Fig:S5}, the $TS$ distributions do not obey the $\chi^2$-distributions given by Wilks' theorem for tests in each single energy bin, while for the joint test the $TS$ distribution is well described by the theorem, despite that the parameter-domain conditions of the theorem are violated.



\begin{thebibliography}{1}
	
	\bibitem{Han2017}
	J.L. Han,
	Observing interstellar and intergalactic magnetic fields, \emph{Annual Review of Astronomy and Astrophysics}
	55:111–57
	(2017).
	
	\bibitem{Subramanian2016}
	K. Subramanian,
	The origin, evolution and signatures of primordial magnetic fields,
	\emph{Rep. Prog. Phys.}
	79, 7
	(2016).
	
	\bibitem{Planck2016}
	The Planck Collaboration,
	Planck 2015 results. XIX. Constraints on primordial magnetic fields,
	\emph{Astron. Astrophys.}
	594, A19
	(2016).
    
    \bibitem{Aharonian1994}
	F.A. Aharonian, P. S. Coppi, H. J. Voelk,
    Very high energy gamma rays from active galactic nuclei: Cascading on the cosmic background radiation fields and the formation of pair halos
    \emph{Astrophys. J. Lett.}
    423, L5 
    (1994.)
	
	\bibitem{Neronov2010a}
	A. Neronov and I. Vovk,
	Evidence for strong extragalactic magnetic fields from Fermi observations of TeV blazars,
	\emph{Science}
	328, 5974
	(2010).
	
	\bibitem{Murase2008}
	K. Murase, K. Takahashi, S. Inoue, K. Ichiki, and S. Nagataki,
	Probing intergalactic magnetic fields in the GLAST era through pair echo emission from TeV blazars,
	\emph{Astrophys. J.}
	686, L67
	(2008).
	
	\bibitem{Essey2011}
	W. Essey, S. Ando, and A. Kusenko,
	Determination of intergalactic magnetic fields from gamma ray data,
	\emph{Astroparticle Physics}
	35, 135
	(2011).
	
	\bibitem{Arlen2014}
	T. C. Arlen, V. V. Vassiliev, T. Weisgarber, S. P. Wakely, and S. Y. Shafi,
	Intergalactic magnetic fields and gamma-ray observations of extreme TeV blazars,
	\emph{Astrophys. J.}
	796, 18
	(2014).
	
	\bibitem{Tanaka2014}
	Y. T. Tanaka, \L. Stawarz, J. Finke, C. C. Cheung, C. D. Dermer, J. Kataoka, A. Bamba, G. Dubus, M. De Naurois, S. J. Wagner, Y. Fukazawa, and D. J. Thompson,
	Extreme blazars studied with Fermi-LAT and Suzaku: 1ES 0347-121 and blazar candidate HESS J1943+213,
	\emph{Astrophys. J.}
	787, 155
	(2014).
	
	\bibitem{Neronov2011}
	A. Neronov, D. V. Semikoz, P. G. Tinyakov, and I. I. Tkachev,
	No evidence for gamma-ray halos around active galactic nuclei resulting from intergalactic magnetic fields,
	\emph{Astron. Astrophys.}
	526, A90
	(2011).
	
	\bibitem{Ackermann2013}
	M. Ackermann, M. Ajello, A. Allafort, K. Asano, W. B. Atwood, L. Baldini, J. Ballet, G. Barbiellini, D. Bastieri, K. Bechtol, \emph{et al.},
	Determination of the point-spread function for the Fermi Large Area Telescope from on-orbit data and limits on pair halos of active galactic nuclei,
	\emph{Astrophys. J.}
	765, 54
	(2013).
	
	\bibitem{Chen2015a}
	W. Chen, J. H. Buckley, and F. Ferrer,
	Search for GeV $\gamma$-ray pair halos around low redshift blazars,
	\emph{Phys. Rev. Lett.}
	115, 211103
	(2015).
	
	\bibitem{Tashiro2013}
	H. Tashiro and T. Vachaspati,
	Cosmological magnetic field correlators from blazar induced cascade,
	\emph{Phys. Rev. D}
	87, 123527
	(2013).
	
	\bibitem{Tashiro2014}
	H. Tashiro, W. Chen, F. Ferrer, and T. Vachaspati,
	Search for CP violating signature of intergalactic magnetic helicity in the gamma-ray sky,
	\emph{Mon. Not. R. Astron. Soc.}
	445, L41-L45
	(2014).
	
	\bibitem{Tashiro2015}
	H. Tashiro and T. Vachaspati,
	Parity-odd correlators of diffuse gamma-rays and intergalactic magnetic fields,
	\emph{Mon. Not. R. Astron. Soc.}
	448, 299
	(2015).
	
	\bibitem{Chen2015b}
	W. Chen, B. D. Chowdhury, F. Ferrer, H. Tashiro, and T. Vachaspati,
	Intergalactic magnetic field spectra from diffuse gamma-rays,
	\emph{Mon. Not. R. Astron. Soc.}
	450, 3371–3380
	(2015).
	
	\bibitem{Broderick2012}
	A. E. Broderick, P. Chang, and C. Pfrommer,
	The cosmological impact of luminous TeV blazars. I. Implications of plasma instabilities for the intergalactic magnetic field and extragalactic gamma-ray background,
	\emph{Astrophys. J.}
	752, 22
	(2012).
	
	\bibitem{Schlickeiser2012}
	R. Schlickeiser, D. Ibscher, and M. Supsar,
	Plasma effects on fast pair beams in cosmic voids,
	\emph{Astrophys. J.}
	758, 102
	(2012).
	
	\bibitem{Miniati2013}
	F. Miniati and A. Elyiv,
	Relaxation of blazar-induced pair beams in cosmic voids,
	\emph{Astrophys. J.}
	770, 54
	(2013).
	
	\bibitem{Sironi2014}
	L. Sironi and D. Giannios,
	Relativistic pair beams from TeV blazars: A source of reprocessed GeV emission rather than intergalactic heating,
	\emph{Astrophys. J.}
	787, 49
	(2014).
	
	\bibitem{Kempf2016}
	A. Kempf, P. Kilian, and F. Spanier,
	Energy loss in intergalactic pair beams: Particle-in-cell simulation,
	\emph{Astron. Astrophys.}
	585, A132
	(2016).
	
	\bibitem{Neronov2010b}
	A. Neronov, D. Semikoz, M. Kachelriess, S.
	Ostapchenko, and A. Elyiv,
	Degree-scale GeV ``jets" from active and dead TeV blazars,
	\emph{Astrophys. J. Lett.}
	719, L130
	(2010).
	
	\bibitem{MOJAVE}
	M. L. Lister, M. F. Aller, H. D. Aller, M. A. Hodge, D. C. Homan, Y. Y. Kovalev, A. B. Pushkarev, and T. Savolainen,
	MOJAVE. XV. VLBA 15 GHz total intensity and polarization maps of 437 parsec-scale AGN jets from 1996 to 2017,
	\emph{Astrophys. J. Suppl. Ser.}
	232, 12
	(2018).
	
	\bibitem{note1}
	The AGN SED class and redshift can be obtained from the MOJAVE data archive: http://www.physics.purdue.\\edu/astro/MOJAVE/allsources.html.
	
	\bibitem{Lister2013}
	M. L. Lister, M. F. Aller, H. D. Aller, D. C. Homan, K. I. Kellermann, Y. Y. Kovalev, A. B. Pushkarev, J. L. Richards, E. Ros, and T. Savolainen,
	MOJAVE. X. Parsec-scale jet orientation variations and superluminal motion in AGN,
	\emph{Astrophys. J.}
	146, 5
	(2013).
	
	\bibitem{Lister2016}
	M. L. Lister, M. F. Aller, H. D. Aller, D. C. Homan, K. I. Kellermann, Y. Y. Kovalev, A. B. Pushkarev, J. L. Richards, E. Ros, and T. Savolainen,
	MOJAVE XIII. Parsec-Scale AGN jet kinematics analysis based on 19 years of VLBA observations at 15 GHz,
	\emph{Astrophys. J.}
	152, 12
	(2016).
	
	\bibitem{TeVCat}
	S. P. Wakely and D. Horan,
	TeVCat: an online catalog for very high energy gamma-ray astronomy,
	\emph{International Cosmic Ray Conference}
	3, 1341-1344
	(2008).
	
	\bibitem{2FHL}
	M. Ackermann, M. Ajello, W. B. Atwood, L. Baldini, J. Ballet, G. Barbiellini, D. Bastieri, J. Becerra Gonzalez, R. Bellazzini, E. Bissaldi, \emph{et al.},
	2FHL: The second catalog of hard Fermi-LAT sources,
	\emph{Astrophys. J. Suppl. Ser.}
	222, 1
	(2016).
	
	\bibitem{FermiSSC}
	https://fermi.gsfc.nasa.gov/ssc/.
	
	\bibitem{Fermi_performance}
	https://www.slac.stanford.edu/exp/glast/groups/canda\\/lat\_Performance.htm.
	
	\bibitem{Pushkarev2017}
	A. B. Pushkarev, Y. Y. Kovalev, M. L. Lister, and T. Savolainen,
	MOJAVE XIV. Shapes and opening angles of AGN jets,
	\emph{Mon. Not. R. Astron. Soc.}
	468:4, 4992-5003
	(2017).
	
	\bibitem{Neronov2009}
	A. Neronov and D. V. Semikoz,
	Sensitivity of $\gamma$-ray telescopes for detection of magnetic fields in the intergalactic medium,
	\emph{Phys. Rev. D}
	80, 123012
	(2009).
	
	\bibitem{Protassov2002}
	R. Protassov, D. A. van Dyk, A. Connors, V. L. Kashyap, and A. Siemiginowska,
	Statistics, handle with care: Detecting multiple model components with the likelihood ratio test,
	\emph{Astrophys. J.}
	571, 545
	(2002).
	
	\bibitem{Long2015}
	A. J. Long and T. Vachaspati,
	Morphology of blazar-induced gamma ray halos due to a helical intergalactic magnetic field,
	\emph{Journal of Cosmology and Astroparticle Physics}
	Vol. 2015
	(2015).
	
	\bibitem{Batista2016}
	R. A. Batista, A. Saveliev, G. Sigl, and T. Vachaspati,
	Probing intergalactic magnetic fields with simulations of electromagnetic cascades,
	\emph{Phys. Rev. D}
	94:8, 083005
	(2016).
	
	\bibitem{Duplessis2017}
	F. Duplessis and T. Vachaspati,
	Probing stochastic inter-galactic magnetic fields using blazar-induced gamma ray halo morphology,
	\emph{Journal of Cosmology and Astroparticle Physics}
	Vol. 2017
	(2017).
    
    \bibitem{Marscher2008}
    A. P. Marscher, S. G. Jorstad, F. D. D'Arcangelo, P. S. Smith, G. G. Williams, V. M. Larionov, H. Oh, A. R. Olmstead, M. F. Aller, H. D. Aller, I. M. McHardy, A. L\"{a}hteenm\"{a}ki, M. Tornikoski, E. Valtaoja, V. A. Hagen-Thorn, E. N. Kopatskaya, W. K. Gear, G. Tosti, O. Kurtanidze, M. Nikolashvili, L. Sigua, H. R. Miller, and W. T. Ryle,
    The inner jet of an active galactic nucleus as revealed by a radio-to-$\gamma$-ray outburst,
    \emph{Nature}
    452, 966-969
    (2008).

	\bibitem{CTA_performance}
	This research has made use of the CTA instrument response functions provided by the CTA Consortium and Observatory, see http://www.cta-observatory.org/\\science/cta-performance/ (version prod3b-v1) for more details.
    
	\bibitem{note2}
	Although conditions of Wilks theorem are not satisfied as we discussed in Section \ref{sec:model and test}, a MC simulation of the likelihood-ratio test statistic is computational hard, especially for a large sample size. Here we use Wilks theorem to calculate the $p$-value in Fig. \ref{Fig:6} to speed up the computation because we would only like to roughly display the trend of the significance as a function of the count rate rather than accurately calculate the p-value for each test.   
	
	\bibitem{Franceschini2008}
	A. Franceschini, G. Rodighiero, and M. Vaccari,
	Extragalactic optical-infrared background radiation, its time evolution and the cosmic photon-photon opacity,
	\emph{Astron. Astrophys.}
	487, 3
	(2008).
	
	\bibitem{FermiEBL2012}
	M. Ackermann, M. Ajello, A. Allafort, P. Schady, L. Baldini, J. Ballet, G. Barbiellini, D. Bastieri, R. Bellazzini, R. D. Blandford, E. D. Bloom, A. W. Borgland, E. Bottacini, A. Bouvier, J. Bregeon, M. Brigida, P. Bruel, R. Buehler, \emph{et al.},
	The Imprint of the Extragalactic Background Light in the Gamma-Ray Spectra of Blazars,
	\emph{Science}
	338, 6111
	(2012).
	
	\bibitem{note3}
	Computational times are calculated based on a platform with Intel Xeon CPU E3-1505 v5 and NVIDIA Quadro GPU M2000M.
	
	\bibitem{cuda9.2}
	https://developer.nvidia.com/cuda-toolkit.
	
\end{thebibliography}
\end{document}